\documentclass[a4paper,11pt]{article}
\pdfoutput=1 

\usepackage{jinstpub} 
\usepackage{verbatim}
\usepackage{lineno}
\usepackage{scrextend}

\title{\boldmath Nucleation efficiency of nuclear recoils in bubble chambers}

\author{D. Durnford}
\author{and M.-C. Piro}

\affiliation{Department of Physics, University of Alberta, Edmonton, Alberta, T6G 2R3, Canada}

\emailAdd{ddurnfor@ualberta.ca, mariecci@ualberta.ca}

\abstract{Bubble chambers using liquid xenon (and liquid argon) have been operated (resp. planned) by the Scintillating Bubble Chamber (SBC) collaboration for GeV-scale dark matter searches and CE$\nu$NS from reactors. This will require a robust calibration program of the nucleation efficiency of low-energy nuclear recoils in these target media. Such a program has been carried out by the PICO collaboration, which aims to directly detect dark matter using $\mathrm{C_3 F_8}$ bubble chambers. Neutron calibration data from mono-energetic neutron beam and Am-Be source has been collected and analyzed, leading to a global fit of a generic nucleation efficiency model for carbon and fluorine recoils, at thermodynamic thresholds of $2.45$ and $3.29\,\mathrm{keV}$. Fitting the many-dimensional model to the data ($34$ free parameters) is a non-trivial computational challenge, addressed with a custom Markov Chain Monte Carlo approach, which will be presented. Parametric MC studies undertaken to validate this methodology are also discussed. This fit paradigm demonstrated for the PICO calibration will be applied to existing and future scintillating bubble chamber calibration data.}

\collaboration[c]{on behalf of SBC and PICO collaborations}

\proceeding{LIDINE 2021: LIght Detection In Noble Elements\\
  Sep.14-17, 2021\\
  University of California San Diego}

\begin{document}
\maketitle
\flushbottom

\section{Introduction}
\label{sec:intro}
The search for dark matter (DM) and coherent elastic neutrino nucleus scattering (CE$\nu$NS) through the observation of low-energy nuclear recoils requires detector technologies designed to be sensitive to sub-keV thresholds while having a high power of background rejection \cite{DM_CEvNS}. The Scintillating Bubble Chamber (SBC) experiment uses novel low-background detectors with noble liquids and a 100 eV energy threshold, aimed at detecting low-mass (0.7-7 GeV/c$^2$) DM interactions and CE$\nu$NS from reactor neutrinos \cite{ICHEP, CEvNS_SBC}. The advantage of such technology is that it combines the electron recoil discrimination of bubble chambers and energy reconstruction from scintillation light. The first demonstration of a scintillating bubble chamber was a successfully operated 30g liquid xenon (LXe) bubble chamber \cite{sbc_first}, establishing noble-liquid bubble chambers as a promising new technology to reach these physics goals. However, understanding the detector response and nucleation efficiency to low-energy nuclear recoils is necessary for the interpretation of experimental results. 

Bubble chambers use super-heated liquids to search for rare, low-energy nuclear recoils. The threshold energy is set by the pressure/temperature conditions of the target fluid. Particle interactions with energy above threshold induce a small volume of liquid to rapidly transition into a vapor state. This nucleation can be observed optically with cameras, acoustically with piezo-electric transducers, or barometrically. The process of particle-induced bubble nucleation is typically described by Seitz’s “hot spike” model \cite{Seitz}. However, it is well known from the PICO collaboration that the threshold for nucleation deviates from the predicted Seitz threshold \cite{pico2L}. A new approach has been developed to determine the nucleation efficiency of nuclear recoils in super-heated liquids \cite{jin_thesis} with dedicated neutron calibrations carried out by the PICO collaboration. In this paper, the preliminary results will be presented for both conventional and noble liquids bubble chambers.\footnote{\label{note1}Publication with more details in preparation.}


\section{PICO Data and Simulations}


Dedicated neutron calibrations over the past 6 years at various thermodynamic thresholds ranging from 2.1 keV to 3.9 keV as summarized in Table \ref{data},  have been performed by the PICO collaboration to measure the nucleation efficiency in super-heated $\mathrm{C_3F_8}$. This includes experiments done with two detectors (PICO-0.1L \cite{pico0.1L} and PICO-2L \cite{pico2L}), with neutron sources (Am-Be and Sb-Be compound sources) as well as neutron beams generated with a TANDEM accelerator at the Universit\'e de Montr\'eal \cite{Tandem}. The proton beam is used to produce mono-energetic neutron with energies of 50, 61, and 97 keV obtained with a vanadium target via the reaction $^{51}$V(p,n)$^{51}$Cr \cite{vanadium}. Fluorine cross-section resonances at 50 and 97 keV, and an anti-resonance 61 keV, break the degeneracy between neutron scattering on carbon and fluorine \cite{pico0.1L}. Neutrons can also scatter multiple times in the detector as they slow down creating multiple bubble events, constraining the nucleation efficiency at several recoil energies simultaneously. Each experiment was simulated using Geant4 \cite{geant4} or MCNPX-Polimi \cite{MCNP} to provide the spectrum of energy depositions on each target atoms species in the corresponding chamber, for all multiplicities. $^{3}$He counters were deployed to measure source strengths and normalize the simulated spectra.
\begin{table}
\protect\begin{centering}
    \begin{tabular}{| c | c | c | c |}
    \hline
    Dataset & Detector & Thresholds (keV) & Multiplicity \\ \hline
    97 keV Beam & PICO-0.1 (2013, 2014) & 3.0, 3.2, 3.6 & 1,2,3+ \\ \hline
    61 keV Beam & PICO-0.1 (2013, 2014) & 2.9, 3.1, 3.6 & 1,2,3+ \\ \hline
    50 keV Beam & PICO-0.1 (2014) & 2.5, 3.5 & 1,2,3+ \\ \hline
    SbBe & PICO-0.1 & 2.1, 2.6, 3.2 & 2,3+ \\ \hline  
    AmBe & PICO-2L & 3.2 & 1,2,3,4,5,6,7+ \\  \hline
    \end{tabular}
\par\end{centering}
\protect\caption{PICO neutron calibration data sets used in this analysis.} 
\label{data}
\end{table}

\section{Analysis and results}
In the absence of an \emph{a prior} physical model for the shape of the nucleation efficiency curve, a piece-wise linear function is used, with $4$ linear segments between $5$ fixed efficiency knots at $0$, $0.2$, $0.5$, $0.8$, and $1$. The recoil energies at which these efficiency knots are reached are the free parameters of the model. There are separate nucleation efficiency curves for carbon and fluorine. To span the range of thermodynamic thresholds in the data, two threshold ``fence-posts'' are defined at $2.45\,\mathrm{keV}$ and $3.29\,\mathrm{keV}$. With most of the data clustered around these energies, a linear scaling to the nearest of the two fence-posts is used to interpolate/extrapolate for different thermodynamic states. In total, this yields a model with $4$ nucleation efficiency curves with $20$ free parameters. Additionally, nuisance parameters are added to account for systematics for each of the experiments, including uncertainties on beam fluxes and source strengths, geometry, and pressure/temperature conditions for a total of $34$ parameters.\footref{note1} The associated likelihood function for this model, with contributions summed over each experiment $i$ and events with $j$ bubbles, is given as

\begin{equation}
\label{eq:LL}
\log \mathscr{L} = \sum_{i}\sum_{j}\left(-\nu_{i,j}(\{x_{n,p}\},\{s_k\})+k_{i,j}+ k_{i,j}\log\left(\frac{\nu_{i,j}(\{x_{n,p}\},\{s_k\})}{k_{i,j}}\right)\right) - \sum_{k}\frac{s_k^2}{2}.
\end{equation}

Here $k_{i,j}$ is the number of bubbles observed in each case, and $\nu_{i,j}$ is the expected number of bubbles for a given input set of parameters defining a set of efficiency curves ${x_(n,p)}$, which are then applied to the simulated spectra. The nuisance parameters are included in the final term, with $s_k$ being a scaling factor for the expected standard deviation of each nuisance parameter.

\begin{figure} 
\center
\includegraphics[width=0.7\textwidth]{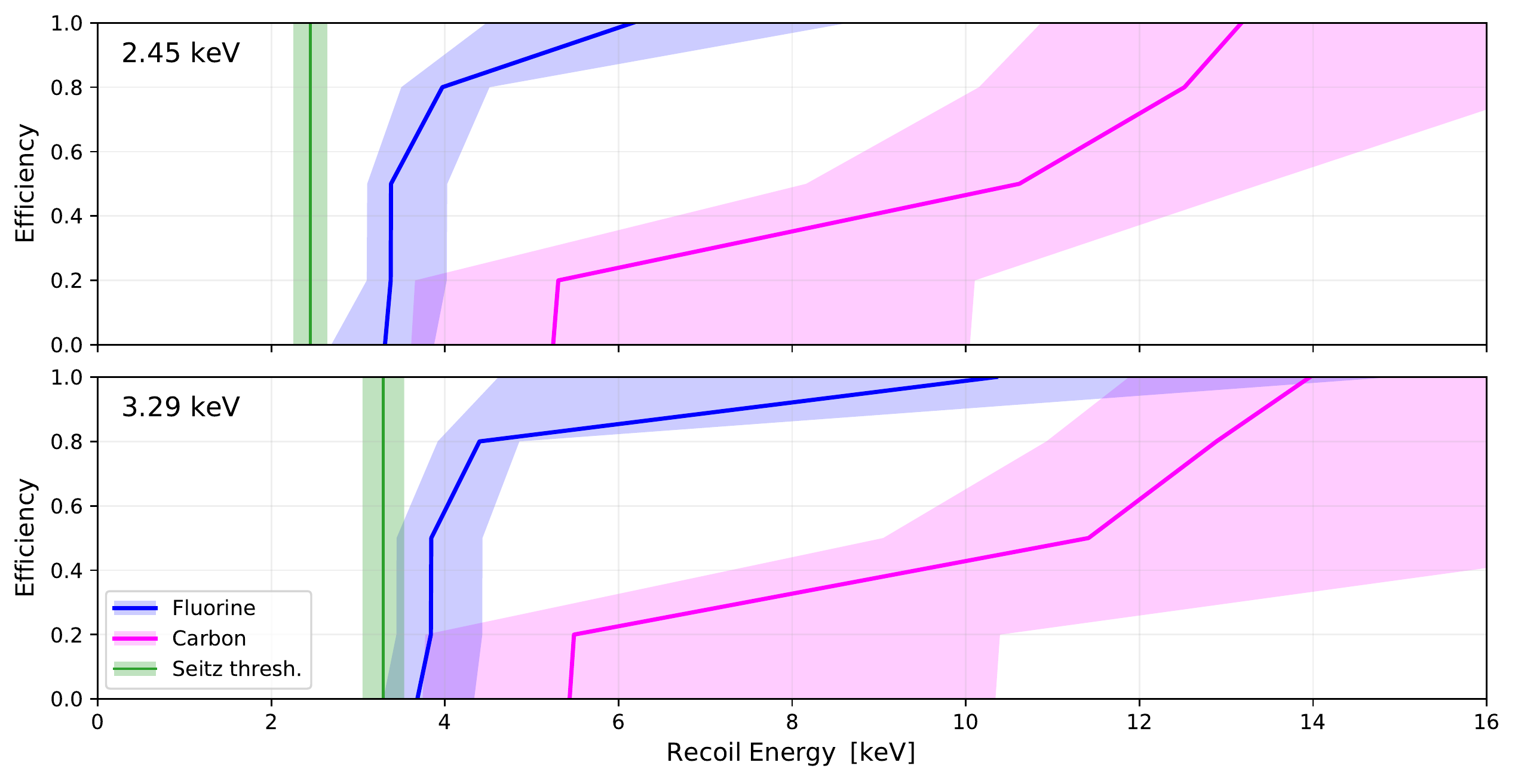}
\caption{Nucleation efficiency curves obtained for the PICO calibration data, showing the best fit curves and $1\,\sigma$ error bands for fluorine and carbon (blue and magenta respectively), and for both thermodynamic thresholds ($2.45\,\mathrm{keV}$ above, $3.29\,\mathrm{keV}$ below) indicated by the green bands.}
\label{fig:result}
\end{figure}

To fit this high-dimensional model, a Markov Chain Monte Carlo (MCMC) \cite{emcee} is a natural choice of optimization strategy, but for which exploring the likelihood function is still computationally expensive. A modified approach was developed \cite{jin_thesis} wherein the MCMC is run for a small number of steps ($5-10$), after which the random walkers are stopped and the cumulative samples are binned in each parameter. The maximum likelihood samples in each bin are then used as starting points for a new set of MCMC walkers. This process is repeated iteratively until convergence is reached. \footref{note1} This forces the MCMC to preferentially explore the boundary of the likelihood function, leading to more efficient computation of the maximum-likelihood point and uncertainty bands (assuming a roughly Gaussian likelihood \cite{cowan}). Convergence was considered to be achieved when the maximum likelihood and span of samples within $1\,\sigma$ of the maximum do not deviate by more than $0.1\%$ for $25$ consecutive MCMC iterations. The results of the fit to the data are shown in figure \ref{fig:result}. It is evident that the nucleation curves deviate significantly from their respective Seitz thresholds, more so for carbon than for fluorine (as expected, since $dE/dx$ will be lower for lighter atoms). This is consistent with the original results obtained with this calibration data \cite{pico60}.

To further study the efficacy of this empirical model for nucleation efficiency and fit-strategy, a parametric Monte Carlo study was carried out. $25$ simulated data sets were constructed using the best-fit model to the data, and fit with the exact same fitting procedure. The first conclusion from this study is confirmation that the convergence criteria used is sufficient. Another result is the ability to interpret the $\chi^2$ value of the best-fit model ($\chi^2 = -2 \times \log \mathscr{L}$) as a goodness-of-fit value. This is non-trivial as the number of degrees of freedom of the fit is confounded by the correlations between many of the parameters. Using the distribution of $\chi^2$ for the MC data sets, the effective number of degrees of freedom of the fit was found to be $46$, giving the fit to the real data a one-sided p-value of $0.189$, proving that the data is reasonably represented by the empirical model used.

Finally, the MC study can be used to check for systematic biases in the model/fit paradigm. Comparing the average MC fit result to the best-fit to the data (the MC truth model) reveals small but persistent offsets for some of the $20$ parameters of interest. However, the fit bias for a given parameter $\theta_i$ may not be a constant offset, but rather a function of that parameter $\theta_i$ or any other $\theta_{j,j \neq i}$. Therefore, a bias study only carried out for MC data generated from one input model only measures the fit bias of each parameter at one point in the parameter space. To better characterize the fit bias, an additional $25$ MC data sets were generated from efficiency curves drawn randomly from within the $1\,\sigma$ band of the real data result, and fit using the same procedure. The results were used to measure the fit bias in the $20$ parameters of interest as a function of all $20$ parameters, fit with a first-order polynomial. This yields $400$ constraints of the form $\theta_{i_\mathrm{fit}} = \theta_{i_\mathrm{true}} + B_{ij}(\theta_{j_\mathrm{true}})$ where $B_{ij}$ is the bias in parameter $i$ as a function of parameter $j$. This was used to construct a likelihood function for the true values of the parameters:

\begin{equation}
    \log \mathscr{L} \left( \left \{\theta_{\mathrm{fit}} \right \} | \left \{ \theta_{\mathrm{true}} \right \} \right) = \sum_j \sum_i \log P \left(\theta_{i_\mathrm{fit}} = \theta_{i_\mathrm{true}} + B_{ij}\left(\theta_{j_\mathrm{true}}\right)\right).
\end{equation}

\begin{figure}[htbp]
\centering 
\includegraphics[width=.47\textwidth]{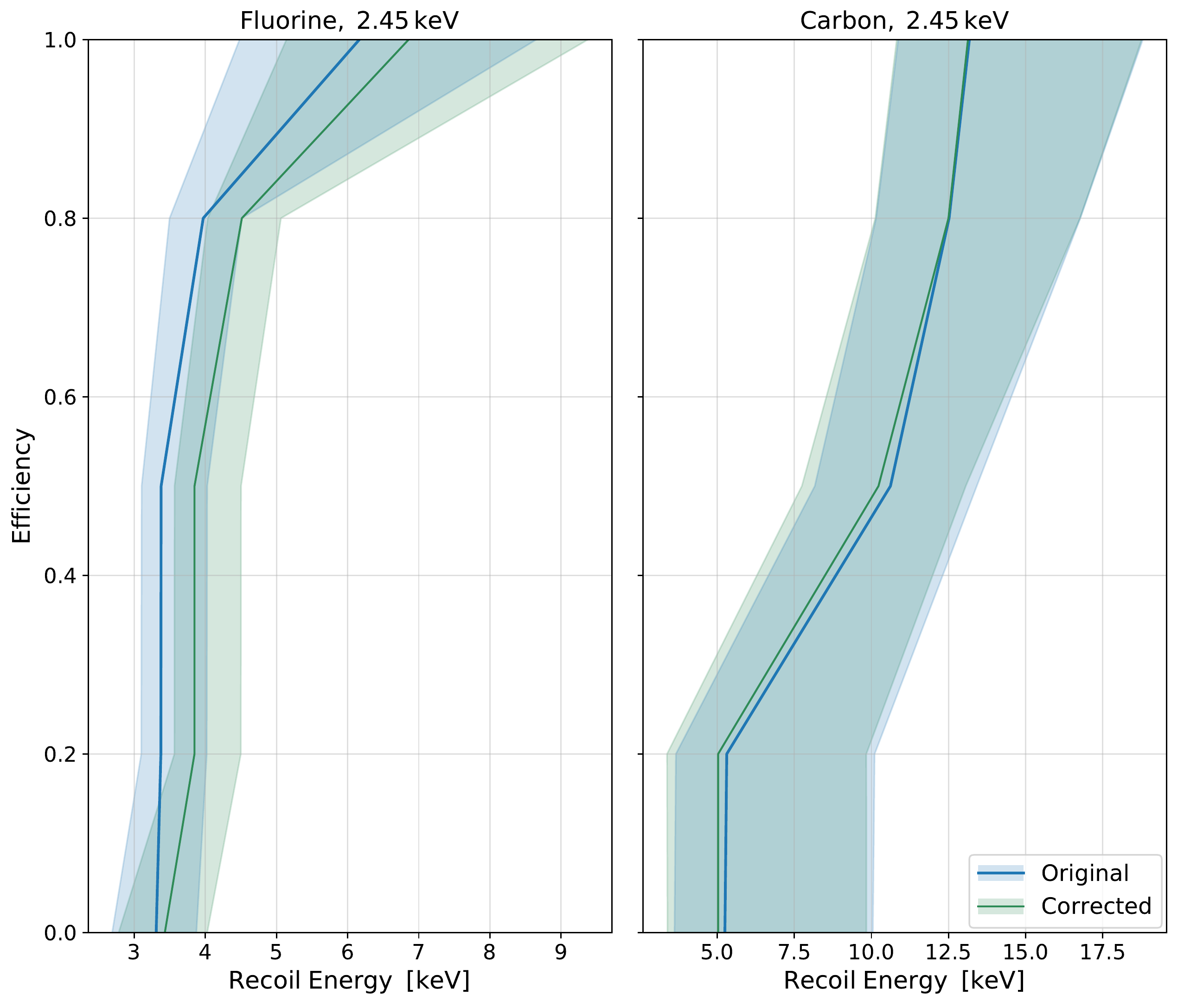}
\includegraphics[width=.47\textwidth]{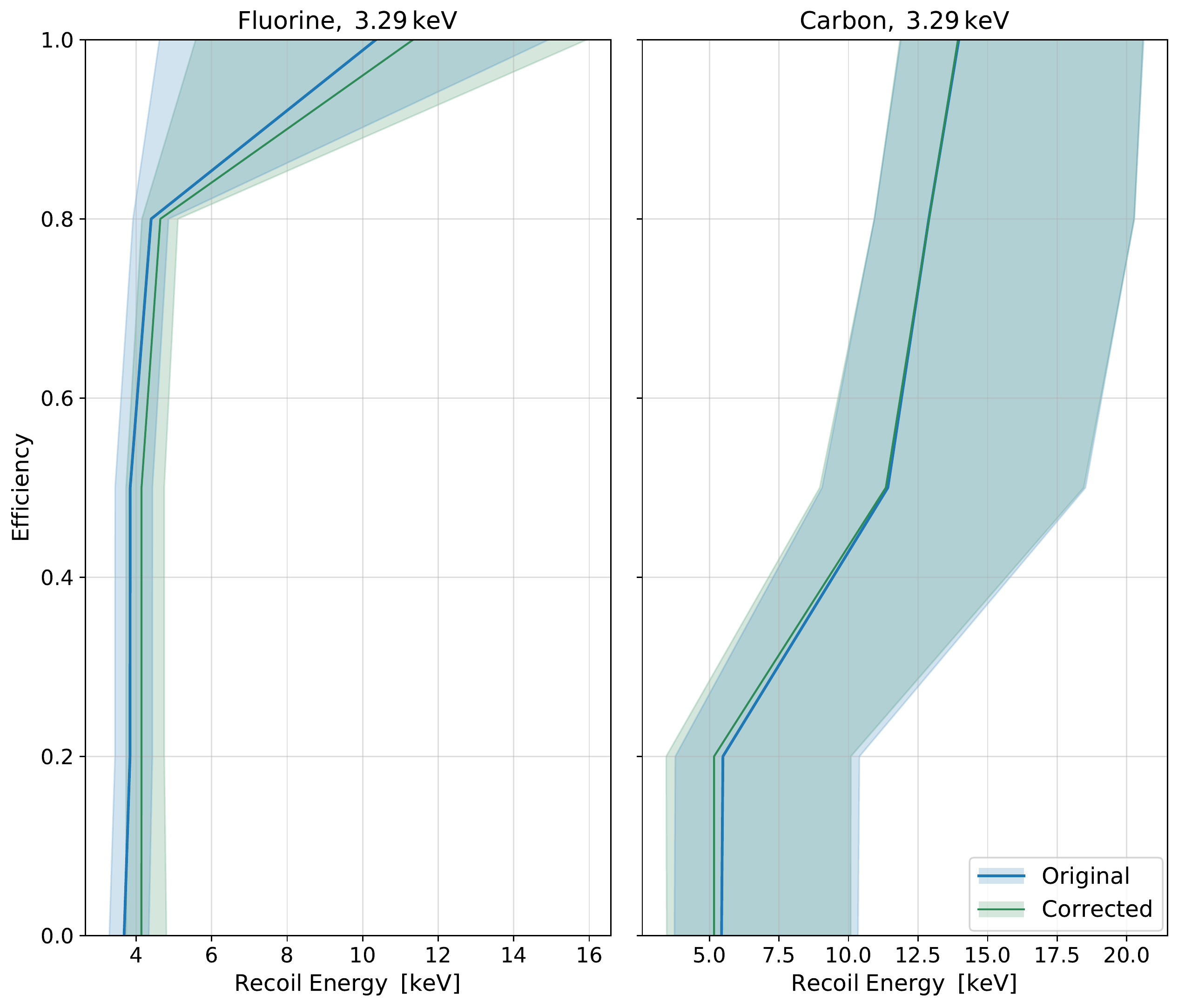}
\caption{\label{fig:bias_corr} Comparison between the original nucleation efficiency curves for PICO (blue) and the bias-corrected results (green), for both thermodynamic thresholds ($2.45\,\mathrm{keV}$ left, $3.29\,\mathrm{keV}$ right).}
\end{figure}

Maximizing this yields a bias-corrected fit result, which is shown alongside the original in figure \ref{fig:bias_corr}. Despite this being a low-powered study, it demonstrates the robustness of the model and fitting approach used and that any systematic biases in this approach are not significant.\footref{note1}

\section{Measurement with a LXe bubble chamber}
\begin{figure} 
\center
\includegraphics[width=0.7\textwidth]{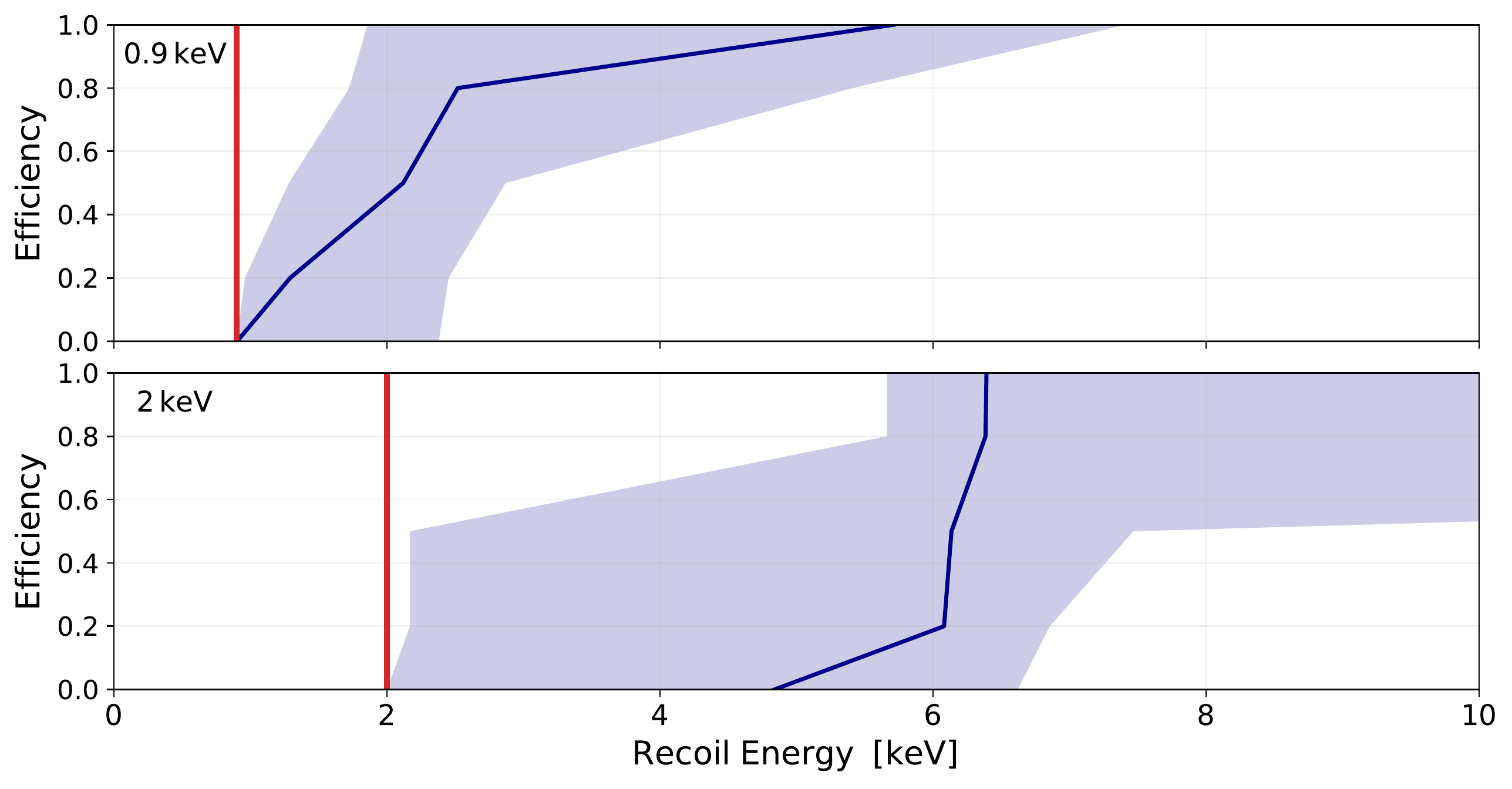}
\caption{Preliminary nucleation efficiency results in LXe (best-fit and $1\,\sigma$ error in blue), for thermodynamic thresholds of $0.9\,\mathrm{keV}$ and $2\,\mathrm{keV}$ (red).}
\label{fig:xe_nr}
\end{figure}

The techniques developed and applied to the PICO calibration data are directly applicable to other bubble chamber experiments such as the SBC collaboration's LXe prototype chamber \cite{sbc_first}. Nuclear recoil calibration data was taken with a $\mathrm{^{252}Cf}$ neutron source and various photo-neutron sources. To date, a Geant4 simulation \cite{geant4} has been produced describing the $\mathrm{^{252}Cf}$ experiment, allowing for an analysis with this data, taken at thermodynamic thresholds of $0.9$, $1.19$, $1.89$, and $2.06\,\mathrm{keV}$. The scintillation signal was used to remove the background signal (the scintillation yield is higher for electronic recoils). As with the PICO analysis, a piece-wise linear model is used for the nucleation efficiency curves, with thermodynamic threshold fence-posts at $0.9$ and $2\,\mathrm{keV}$, and for a single target species. The same interpolation scheme is used for data sets with different thermodynamic states, and nuisance parameters are included to incorporate experimental systematics. The same iterative MCMC approach \cite{jin_thesis} and convergence criteria are used, with the resulting fit shown in figure \ref{fig:xe_nr}. As with the $\mathrm{C_3F_8}$ result, these experimental nucleation efficiency curves from LXe deviate significantly from their theoretical Seitz thresholds, albeit with significant uncertainty. Although this initial analysis is promising in terms of demonstrating sensitivity to $\sim 1\,\mathrm{keV}$ recoils, future work will incorporate the photo-neutron calibration data to better constrain the nucleation efficiency of LXe bubble chambers.

\section{Conclusion}
A method to extract the nucleation efficiency from neutron calibrations has been developed and applied to PICO data, yielding nucleation efficiency curves for $\mathrm{C_3F_8}$ compatible with previously published results \cite{pico60}. This includes a flexible model for nucleation efficiency that is dependent on thermodynamic state and target atom species, as well as a novel MCMC optimization technique to overcome computational challenges. The robustness of this approach has been tested with simulated data sets, demonstrating that any systematic biases are small and correctable, and that the model adequately represents the PICO data. This method can be applied to other bubble chamber target medium, with preliminary results shown for LXe, and will be used by the SBC collaboration for future liquid noble bubble chamber experiments.

\acknowledgments

This research was undertaken, in part, thanks to funding from the Arthur B. McDonald Canadian Astroparticle Physics Research Institute and the Natural Sciences and Engineering Research Council of Canada. We gratefully acknowledge the support of Compute Canada for providing the computing resources required to undertake this work. We also acknowledge the SBC and PICO collaborations.

\end{document}